\documentclass[10pt,pra,aps,showpacs,twocolumn,unsortedaddress]{revtex4-1}
\usepackage{graphicx,bm}
\usepackage{amsmath}
\usepackage{amssymb}
\usepackage{hyperref}
\usepackage[usenames]{color}

\newcommand{\bs}{\boldsymbol}

\newcommand{\lc}{\epsilon} %Levi-Civita symbol

\newcommand{\ve}{\varepsilon}

\DeclareMathAlphabet\mathbfcal{OMS}{cmsy}{b}{n} %define \mathbfcal

\def\be#1\ee{\begin{equation}#1\end{equation}}
\def\ba#1\ea{\begin{align}#1\end{align}}

\begin{document}

\title{Omnidirectional spin Hall effect in a Weyl spin-orbit coupled  atomic gas}
\author{J. Armaitis}
\email{jogundas.armaitis@tfai.vu.lt}

\author{J. Ruseckas}
\email{julius.ruseckas@tfai.vu.lt}

\author{G. Juzeli\={u}nas}
\email{gediminas.juzeliunas@tfai.vu.lt}
\affiliation{Institute of Theoretical Physics and Astronomy,
Vilnius University, 
Saul\.{e}tekio Ave.~3, LT-10222 Vilnius, Lithuania}

\date{\today}

\begin{abstract}
We show that in the presence of a three-dimensional (Weyl) spin-orbit coupling,
a transverse spin current is generated in response to either a constant
spin-independent force or a time-dependent Zeeman field in an arbitrary
direction. This effect is the non-Abelian counterpart of the universal
intrinsic spin Hall effect  characteristic to the two-dimensional Rashba 
spin-orbit coupling.
We quantify the strength of  such an omnidirectional spin Hall effect by
calculating the corresponding conductivity for fermions and non-condensed
bosons. The absence of any kind of disorder in ultracold-atom systems makes the
observation of this effect viable.
\end{abstract}

% 03.65.Vf  Phases: geometric; dynamic or topological
% 03.75.Ss	Degenerate Fermi gases
% 05.30.Jp 	Boson systems (for static and dynamic properties of Bose-Einstein
% condensates, see 03.75.Hh and 03.75.Kk; see also 67.10.Ba Boson degeneracy in
% quantum fluids)
% 85.75.-d	Magnetoelectronics; spintronics: devices exploiting spin polarized
% transport or integrated magnetic fields

\maketitle

\section{Introduction\label{sec:Introduction}} 

Gauge theories and related geometrical concepts play a prominent role in the
description of physics at a wide range of length scales covering all
fundamental interactions \cite{Frankel04}.  On the other hand, when it comes to
effective models, many quantum-mechanical systems with adiabatically varying
parameters are naturally described in terms of Abelian gauge theories
\cite{Berry84,Shapere89,Bohm03}.  This geometric approach based on the Berry
phase has paved the way to a multitude of both theoretical and experimental
developments covering molecular  \cite{Shapere89,Bohm03,Mead92}, solid-state
\cite{Loss90,Levy90,Chandrasekhar91,Mailly93,Neubauer09,Li13,Hasan10}, photonic
\cite{Bliokh08,Hafezi11,Kitagawa12Ncom,Rechtsman13,Hafezi14,Tzuang14,Buljan15,Skirlo15,Ozawa16},
mechanical \cite{Huber15, Salerno16} and electric \cite{Simon15,Albert15}
systems.  Although  the corresponding non-Abelian gauge structure in the
presence of degenerate quantum states has been noticed  promptly after the
discovery of the Berry phase \cite{Wilczek84}, a set of experimentally observed
signatures of the non-Abelian geometrical phases remains limited
\cite{Zee88,Zwanziger90}. 

Ultracold-atom experiments have been recently gaining tools uniquely suited to
address this elusive non-Abelian gauge structure using the internal states of
the atom
\cite{Lin09a, Lin09b, Aidelsburger11, Struck12, Aidelsburger13, Ketterle13,
Kennedy13, Armaitis13, Choi13, Jotzu14, Aidelsburger14, Atala14, Flaschner16,
Li16, Spielman16Science, Fallani16Science}.
In particular, engineering various species of spin-orbit coupling
\cite{Dudarev04, Ruseckas05, Osterloh2005, Juzeliunas08PRA, Stanescu2007,
Vaishnav08PRL, Liu2009, Chuanwei-Zhang10, Campbell2011, Dalibard11, 3dSOCprl,
Xu2013, magSOCprl, Wu13JPB, Brandon13, Li2013, Galitski13-review,
SOCreview14, Zhai14-review}
in ultracold-atom systems has seen rapid advances lately
\cite{Lin2011, Zhang2012, Wang2012, Cheuk2012, Williams2012, LeBlanc2013,
Engels2013, Fu2014, Huang16NP, Meng15, Pan15},
allowing experimental demonstration of, e.g., the phase diagram of spin-orbit
coupled (SOC) bosons \cite{Ji14} and the spin Hall effect \cite{Beeler13}.
However, despite the existence of this novel toolbox, there is a lack of
concrete proposals to unambiguously demonstrate the non-Abelian gauge
structure.

The spin Hall effect (SHE), in which density currents generate transverse spin
currents has already played a prominent role in condensed-matter physics and
has provided an impetus to the field of spintronics \cite{SinovaRev}. The SHE
has been detected experimentally in a wide variety of solid-state materials,
which usualy possess a planar SOC of the Rashba-Dresselhaus type.  In these
solid-state experiments, the SOC plane corresponds to the two-dimensional (2D)
geometry of the sample. Therefore, both the perturbation of the system (applied
voltage) and the resulting spin current are confined to that plane.  On the
other hand, the spin current is polarized in the direction perpendicular to the
SOC plane.  This spin Hall effect is induced by a spin-dependent Berry magnetic
field (Berry curvature) perpendicular to the plane.  Such a magnetic field is
proportional to a single Pauli matrix $\sigma_z$ and hence is Abelian, as it
will be discussed in more details in the next Section.

\begin{figure}[t]
\includegraphics[width=0.49\linewidth]{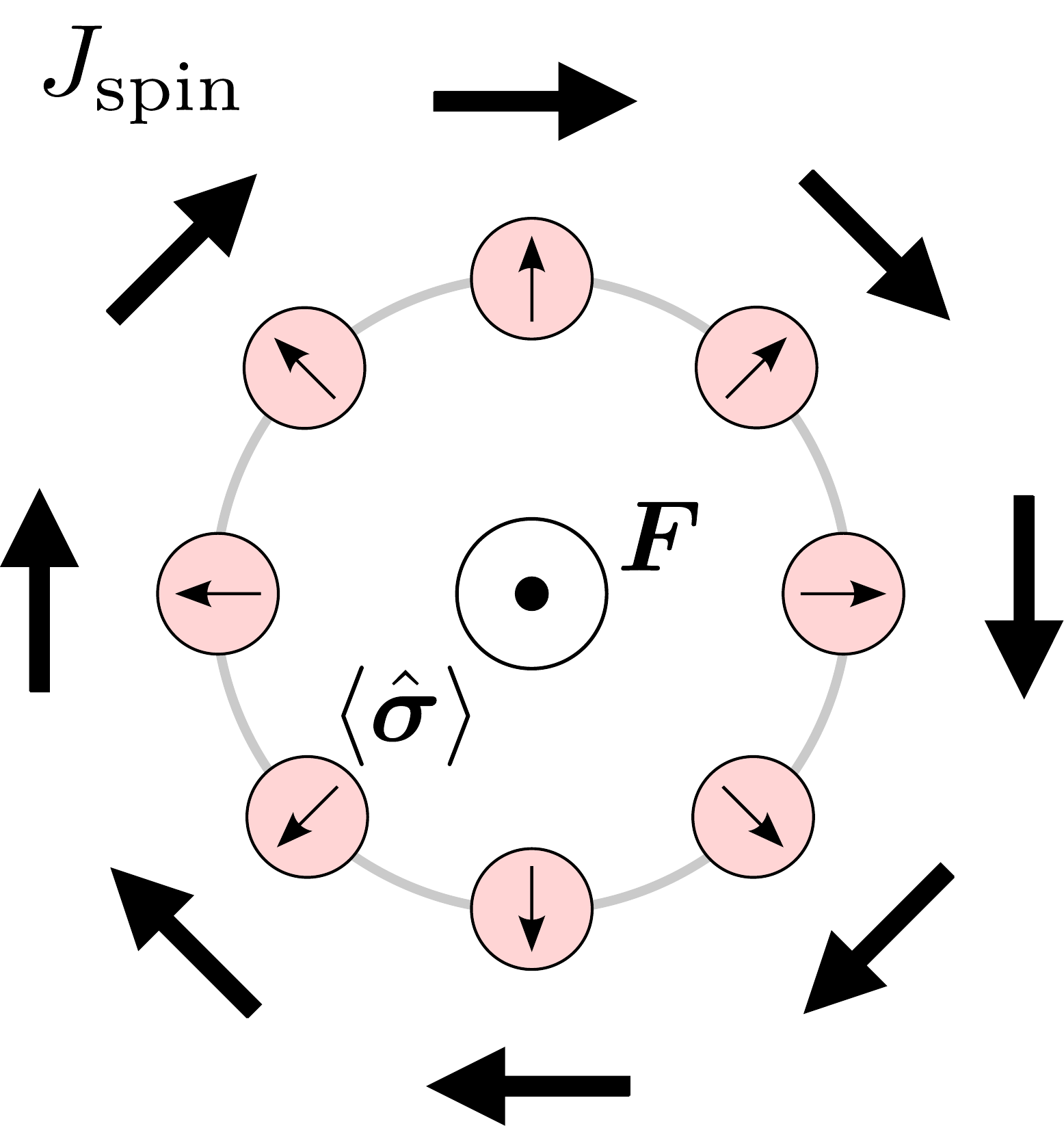}
\caption{In a Weyl spin-orbit coupled system, a spin-independent force
  pointing in an arbitrary direction drives a spin current in the plane
  perpendicular to the force.  Here we show this situation in the momentum
  space.  Particles with spins  $\langle \hat {\bs \sigma} \rangle$ respond to
  the force $\bs F$ by contributing to  the spin current $J_\text{spin}$ which
  runs perpendicular to the force and carries the spin orthogonal to  both the
  current and the force, see Eq.~\eqref{eq:spin-current-T}.  The independence
  of this phenomenon from the driving direction unambiguously demonstrates the
  non-Abelian nature of the underlying dynamics.}
\label{fig:main}
\end{figure}

In the present article we put forward a three-dimensional (3D) version of the
SHE based on the novel possibility to engineer a nonplanar spin-orbit coupling
of the Weyl type for ultracold atoms 
\cite{3dSOCprl,magSOCprl,Wu13JPB,Brandon13,Tokatly16}.
In the proposed setup, the atoms are affected by a 3D Berry magnetic field
which is non-Abelian and induces  a spin-dependent Lorentz-type force  for all
directions of atomic motion.  Perturbing the system  along an arbitrary axis
produces a spin current perpendicular to the  perturbation
(Fig.~\ref{fig:main}).  Such a response is in  a stark contrast to the Abelian
case, where the magnetic field unavoidably has a single well-defined
direction, and the (spin) Hall effect occurs in the plane perpendicular to it.
We will refer to the present effect based on the 3D SOC as an omnidirectional
spin Hall effect which is the non-Abelian counterpart of the universal
intrinsic spin Hall effect  characteristic to the two-dimensional Rashba SOC
\cite{SinovaPRL}. 

In certain  lattice systems
\cite{Burkov11, Ando13, Borisenko2014exp,
Liu2014exp, Xu2015exp, Lu15exp, Soluyanov2015, Lv2015exp, Buljan15PRL},
pairs of Weyl points of opposite topological charges arise governing the topological properties
\cite{Burkov11, Wan11} or  interactions between particles \cite{Gursoy13} in
the so-called Weyl semimetal regime. Since the Weyl points have opposite 
topological charges, they respond to driving in an opposite way, 
and the induced spin-currents cancel.  Here we consider the
SOC of the Weyl type (also known as the Weyl-Rashba SOC) produced by 
manipulating  atomic internal  states  \cite{3dSOCprl,magSOCprl,Wu13JPB,Brandon13}
rather than using a lattice,  so only a single Weyl 
point arises. This is an important feature for generating a nonzero spin 
current in response to a spin-independent force. 

In the present study we do not include the effects due to impurities.  The
impurities play a crucial role in the spin Hall effect physics for electrons in
solids to the extent of preventing  the universal intrinsic spin Hall effect
\cite{Dimitrova2005, Khaetskii2006, SinovaRev}. However, ultracold-atoms are
free from impurity scattering, both magnetic and nonmagnetic, so the spin-Hall
effect is not suppressed in these systems.

Furthermore, interactions between the ultracold-atoms are typically weak
\cite{Leggett2001}, and they can be further minimized by utilizing  the
Feshbach resonances \cite{Chin2010}. We therefore leave the detailed study of
interaction effects \cite{Guan17} for future work. 

The paper is organized as follows. In Section ~\ref{sec:Framework} we define
the  atomic Hamiltonian  with the Weyl SOC  included, and write down the
equations of motion for the spin and center-of-mass degrees of freedom.
Section~\ref{sec:Spin current} explores spin currents in this system, and
presents the possibility to generate a transverse spin current for any
direction of the applied perturbation.  The concluding
Section~\ref{sec:Summary} summarizes the findings and outlines possible future
directions. In Appendix, we discuss in more detail the definition of the spin
current  used in the main text, and consider a relationship between spin
(Stern-Gerlach) projection measurement and and the spin Hall current.

\section{Theoretical framework and non-Abelian dynamics\label{sec:Framework}} 

\subsection{Hamiltonian\label{subsec:Hamiltonian}} 

Let us consider an ensemble of atoms subjected to a Weyl (3D) SOC of a
strength $\chi$.  Individual atoms are described by the Hamiltonian
\be
\hat{H}= (\bs p - \hat{\mathbfcal A})^2/2m +U,
\quad 
\hat{\mathbfcal A} = \chi \hat{\bs \sigma} + \beta \bs e_\beta\,,
\label{eq:H}
\ee
where the Weyl SOC is due to the vector of Pauli matrices $\hat{\bs \sigma} =
\hat \sigma_x \bs e_x + \hat \sigma_y \bs e_y + \hat \sigma_z \bs e_z$ entering
the vector potential $\hat{\mathbfcal A} $.  An extra term $\beta \bs e_\beta$
provides  a spatially uniform spin-independent driving force $\dot{\beta} \bs
e_\beta$ perturbing the atoms  along  a unit vector $e_\beta$, the dot denoting
a time derivative.  Here $U$ is a spin-independent trapping potential. We suppress the identity matrix in the spin space and set
$\hbar=1$ at the outset.  In Eq.~\eqref{eq:H} the bold font specifies a
spatial vector, whereas  the hat indicates an operator acting on the atomic
internal (pseudo-) spin states.  Moreover, $\bs p$ is  a momentum operator and
$m$ is  an atomic mass.  Although for concreteness we consider (pseudo-) spin
1/2 atoms, generalization to a higher-spin system is straightforward and does
not change the qualitative picture.

The Hamiltonian in Eq.~\eqref{eq:H} yields two dispersion branches for an
unperturbed atom ($\beta=0$) affected by  the Weyl SOC: 
\be
\ve_{p\pm} = 
\left(p^2
\pm
2{\chi p}
+\chi ^2
\right) /2m
=\left(p
\pm
\chi
\right)^2 /2m
\,,
\label{dispersion}
\ee
where the lower (upper) sign corresponds to the lower (upper) dispersion branch
in which the spin points along (opposite to) the momentum $\bs p$. In writing
Eq.~(\ref{dispersion}) we have added a constant to place  the minimum of the
lower dispersion branch at the zero energy: $\ve_{\chi -}=0$.

\subsection{Equations of motion\label{subsec:Equations}} 

Defining a velocity operator  for an atom  via the Heisenberg equation 
\be
\hat {\bs v} = -i[\bs r, \hat H] = (\bs p - \hat{\mathbfcal A})/m\,, 
\label{v-definition}
\ee
one can rewrite the Hamiltonian in a concise manner:
$\hat H = m \hat {\bs v}^2/2 + U$, where  $\bs r$ is  a position operator.  The velocity
operator $\hat {\bs v}$ contains a vector potential $\hat{\mathbfcal A}$ which
is an operator acting in the spin space.  Hence $\hat {\bs v}$ obeys the
following non-trivial Heisenberg equation of motion \cite{SOCreview14}
\be
m\dot{\hat {\bs v}} = (\hat {\bs v}\times \hat{\mathbfcal B}
- \hat{\mathbfcal B}\times\hat 
{\bs v})/2 -\bs\nabla 
U
+ \hat{\mathbfcal E}\,,
\label{v-eq}
\ee 
where 
\be
\hat{\mathbfcal E} = -\partial_t \hat{\mathbfcal A}
= -\dot{\beta} \bs e_\beta
\label{eq:BerryE}
\ee
is the strength of the perturbing Berry electric field, and
\be
\hat{\mathbfcal B} = -i\hat{\mathbfcal A}\times
\hat{\mathbfcal A} = 2\chi^2 \hat{\bs \sigma}
\label{eq:B}
\ee
is the strength of the Berry magnetic field. 

The latter magnetic field $\hat{\mathbfcal B}$ is proportional to the spin
operator $\hat{\bs \sigma}$. Hence, it has non-commuting Cartesian components,
showing a non-Abelian character of $\hat{\mathbfcal B}$. This is in contrast to
the usual planar Rashba-type spin-orbit coupling  for which $\hat{\mathbfcal A}
\propto \hat \sigma_x \bs e_x + \hat \sigma_y \bs e_y $,  so the Berry magnetic
field strength $\hat{\mathbfcal B} \propto \hat \sigma_z \bs e_z$ contains
commuting Cartesian components, and is thus Abelian.  Note that the components
of the   spatially uniform Berry magnetic field \eqref{eq:B} can be written in
terms of the field strength
$\hat {\mathcal F}_{ab} = -i[\hat{\mathcal A}_a,\hat{\mathcal A}_b]$
\cite{Wilczek84}, also known as the Yang-Mills curvature \cite{Fujita11}: $\hat
{\mathcal B}_a = \sum_{b,c}\lc_{abc} \hat {\mathcal F}_{bc}/2$.  Genuine
non-Abelian dynamics occurs only in systems where $[\hat {\mathcal F}_{ab},\hat
{\mathcal F}_{cd}]\neq 0$, as discussed in detail in Ref.~\cite{SOCreview14}.
Indeed, our system falls into the non-Abelian-dynamics class, as here $\hat
{\mathcal F}_{ab} = 2\chi^2 \sum_c \lc_{abc} \hat \sigma_c$.

The spin-dependent part of the Hamiltonian \eqref{eq:H} can be represented as
$-\bs M \cdot \hat{\bs \sigma}$, where we have introduced an effective magnetic
field
\be
\bs M = \chi\bs p^{\prime} /m \,,\quad \bs p^{\prime} = \bs p - \beta \bs e_\beta \,,
\label{eq:M}
\ee
with $\bs p^{\prime}$ being a momentum shifted by the spin-independent driving
term $\beta \bs e_\beta$. The spin dynamics follows a Landau-Lifshitz type
\cite{Landau35} equation (LLE) 
\be
\dot{\hat{\bs \sigma}} = -i[\hat{\bs \sigma}, \hat H]= 2\hat{\bs \sigma}\times\bs M\,,
\label{eq:sigma}
\ee
where for our dissipationless cold-atom system we have not added the Gilbert
damping \cite{Gilbert2004} term usually present when describing the
magnetization dynamics in solids. One can now write down the following concise
equation of motion for the velocity in terms of $\bs M$:
\be
m\dot{\hat{\bs v}} = -\dot \beta \bs e_\beta +
2 \chi \bs M \times \hat{\bs \sigma} -\bs \nabla U \,.
\label{eq:v-1}
\ee

Equations \eqref{eq:sigma} and \eqref{eq:v-1} describe the full atomic dynamics
which involves both internal and center-of-mass degrees of freedom.  From now
on we consider a homogeneous system for which $U = 0$. This condition is viable
in a harmonic trap in the local density approximation \cite{Leggett2001} sense,
or, alternatively, in a flat trap \cite{Gaunt13} away from the boundaries of
the trap.  In such a homogeneous case, the momentum $\bs p$ is an integral of
motion: $\dot{\bs p} = 0$. Therefore, dynamics in the spin sector completely
determines the evolution of the system. 

\subsection{Adiabatic approach\label{subsec:Adiabatic}} 

Since the momentum $\bs p$ is conserved, we will henceforth work with momentum
eigenstates and treat $\bs p$ and $\bs M$ as ordinary vectors rather than
operators. The equation of motion for the quantum expectation value
$\langle\hat{\bs\sigma}\rangle$ then has the same form as Eq.~\eqref{eq:sigma}
for the spin operator $\hat{\bs \sigma}/2$.  Hence one arrives at the following
solution for the expectation value to the first order in time derivatives  of
$\bs M$ (see~Refs.~\cite{Bijl11,Huang16} for more details): 
\be
\langle \hat{ \bs \sigma} \rangle _{\pm} = \pm S\left(\frac{\bs M}{M}
  + \frac{1}{2M^3}\bs M \times \dot{\bs M} \right)\,,
\label{eq:sigma-averag}
\ee
where the upper (lower) sign pertains to upper (lower) dispersion branch in
which spin points along (opposite to) the effective magnetic field. Here
\be
M \equiv |\bs M| = \chi |\bs p^{\prime}|/m\,,\quad
\dot{\bs M} = -\chi \dot \beta \bs e_\beta/m\,,
\label{M,M-dot}
\ee
and the normalization factor 
\be
S=
\big(
1 + [\dot \beta 
{m}/{2\chi |\bs p^{\prime} |^ 2}
]^2
\big)^{-1/2}
\simeq 1
\ee
ensures that $\langle\hat{\bs\sigma}\rangle^2_{\pm} = 1$. The condition
$S\simeq1$ defines a range of validity of Eq.~\eqref{eq:sigma-averag}:
\be
\dot \beta 
{m}/{2\chi |\bs p^{\prime} |^ 2} \ll 1.
\label{eq:adiabatic_condition}
\ee

In this adiabatic approach the spin expectation value is determined by the
momentum-dependent effective magnetic field $\bs M$, as well as by the
correction term containing the time-derivative $\dot{\bs M}$ due to the
external force.  In the zero-order adiabatic approximation, the spin follows
the effective magnetic field:
$\langle \hat{ \bs \sigma}^{(0)} \rangle _{\pm} =
\pm \bs M /M = \pm \bs p^{\prime} /p^{\prime}$. The first-order correction
$\langle \hat{ \bs \sigma}^{(1)} \rangle _{\pm} =
\pm \bs M \times \dot{\bs M}/2M^3$  is given by
\be
\langle \hat{ \bs \sigma}^{(1)} \rangle _{\pm}
 = \mp \dot \beta m \bs p^{\prime} \times  \bs e_\beta / 2\chi p^{\prime 3}  
\approx   \mp \dot \beta m  \bs p \times  \bs e_\beta /2\chi p^{3} \,,
\label{eq:MxM-dot}
\ee
where the last relations also assumes small momentum changes: $|\bs p^{\prime}
- \bs p| = \beta \ll p $.  The correction $\langle \hat{ \bs \sigma}^{(1)}
\rangle _{\pm}$ tilts the spin in the direction orthogonal both to the momentum
of the atom and also orthogonal to the driving force  which can point in an
arbitrary direction $\bs e_\beta$.  Hence, this first-order correction term
induces a transverse spin Hall current  to be considered  in detail in the next
Section. The induced spin current in turn provides a direct signature of the
omnidirectional spin Hall effect illustrated in Fig.~\ref{fig:main}.

\section{Spin current\label{sec:Spin current}}

We use an anticommutator-based definition of the spin current tensor
(see~Appendix and Refs.~\cite{Rashba03,Drouhin11,Sherman14} 
for a detailed discussion),
namely,
\be
(J_\text{spin})_i^j =
\frac{1}{2}
\overline{ \langle  \{ \hat v_i,\hat \sigma_j\} \rangle }
=
(\overline{ p^{\prime}_i \langle \hat   \sigma_j \rangle } 
- \chi n_3 \delta_{ij} )/m\,,
\label{eq:J_spin}
\ee
where $n_3$ is a particle density of our 3D system. The subscript $i$ labels
the position-space components of the current defining the flow direction,
whereas the superscript $j$  indicates the spin components  specifying the spin
direction carried by the current.  Angular brackets  signify a quantum average
over the spin degrees of freedom for a fixed momentum $\bs p$  of an individual
atom.  An overline denotes  a subsequent ensemble average, that is, a
statistical average over momentum eigenstates of the equilibrium atomic
ensemble.  Since we are working in the Heisenberg representation, the dynamics
is contained exclusively in the time dependence of the operators.

The atoms in different dispersion branches contribute differently to the spin
current, so it is convenient to rewrite Eq.~(\ref{eq:J_spin}) as  
\be
(J_\text{spin})_i^j =
(\sum_{\eta=\pm}\overline{ p^{\prime}_i \langle\hat\sigma_j \rangle}_{\eta}
- \chi n_3 \delta_{ij} )/m\,,
\label{eq:J_spin-separation}
\ee
where the upper (lower) sign corresponds to atoms in the  upper (lower)
dispersion branch  labelled by the symbol $\eta = \pm$.

\subsection{Equilibrium spin current\label{subsec:equilibrium-spin-current}} 

At equilibrium the external force is absent ($\beta = 0$), so
$\bs p^{\prime}=\bs p $ and
$\langle \hat{ \bs \sigma}\rangle_{\pm}=
\langle \hat{\bs \sigma}^{(0)} \rangle_{\pm} =\pm \bs p/p $.  Since the
momentum distribution is spherically symmetric, the ensemble averaging yields
for atoms in a selected dispersion branch: 
\be
\overline{  p_i p_j / p}_{\pm} =  \delta_{ij} \overline{p_i p_i / p}_{\pm}=
\delta_{ij}\overline{(p)}_{\pm}/3\,.
\ee 
Consequently the spin current (\ref{eq:J_spin-separation}) takes the form 
\be
(J^0_\text{spin})_i^j =
\left[\sum_{\eta=\pm}\eta\overline{(p)}_{\eta}/3 + \chi n_3 \right] \delta_{ij}/m\,.
\label{eq:J^0_spin}
\ee
As can be seen from this expression, the equilibrium   spin current
$(J^0_\text{spin})_i^j$ generally does not vanish in our system.  This is usual
for SOC systems in general \cite{Tokatly08} and has also been considered in the
context of cold atoms in particular \cite{Phuc15}. Note that at equilibrium,
the spin current is longitudinal, i.e., the  spin is polarized along the
Cartesian vector $\bs e_j$ parallel to the  direction $\bs e_i$ of the spin
current. This is reflected by the Kronecker delta function entering
Eq.~\eqref{eq:J^0_spin}.

\subsection{Spin  Hall current\label{subsec:induced-spin-current}}

In what follows, we concentrate on the spin currents brought about by driving.
Specifically, we will consider the difference in the spin currents between the
driven system and the equilibrium situation, namely, the induced spin current 
\be
(\delta J_\text{spin})_i^j =  \overline{p_i \langle \hat\sigma_j ^{(1)} \rangle } /m \,.
\label{eq:delta-J_spin-definition}
\ee
Calling on Eq.~\eqref{eq:MxM-dot} for $\langle \hat{ \bs \sigma}^{(1)}
\rangle_{\pm}$, the induced current takes the form: 
\be
(\delta J_\text{spin})_i^j = -
\frac{\dot \beta}{2\chi} 
\bs e_j \cdot \left( \bs e_i \times   \bs e_\beta \right)
\sum_{\eta=\pm}
\eta 
\overline{ \left( \frac{p_i p_i}{p^3} \right)}_{\eta} \,.
\label{eq:delta-J_spin}
\ee
Using the fact that the momentum distribution is  spherically symmetric, one
arrives at the following result: 
\be
(\delta J_\text{spin})_i^j = \dot\beta  \sigma_{\text{SH}}
\left( \bs e_i \times \bs e_j \right)  \cdot \bs e_\beta
\,,
\label{eq:delta-J_spin-1}
\ee 
where 
\be
\sigma_{\text{SH}} = -
\frac{1}{6\chi}
\sum_{\eta=\pm} \eta \overline{ \left( 1/p \right)}_{\eta} \,
\label{eq:sigma_SHE-result}
\ee
is  the spin Hall conductivity.  For instance, by choosing the driving to point
along the $z$ axis ($\bs e_\beta= \bs e_z$) the spin and its spin current will
be in the xy plane, as in Fig.~\ref{fig:main}.

In this way, in contrast to the equilibrium spin current, the induced spin
current given by Eqs.~(\ref{eq:delta-J_spin-1})--(\ref{eq:sigma_SHE-result}) is
transverse.  Specifically, the spin current flows in the direction $\bs e_i$
which is perpendicular both to the driving force  $\propto e_\beta$ and also to
the spin that the current carries.  This holds for an arbitrarily directed
driving force and thus represents the omnidirectional spin Hall effect.

It is noteworthy that the two dispersion branches provide opposite contribution
to the spin  conductivity in Eq.~(\ref{eq:delta-J_spin-1}). Therefore
$\sigma_{\text{SH}}$ should be larger at low temperatures  when the atoms
populate predominantly the lower dispersion branch. In the following Section we
shall explore this issue in more details.

At this point it is useful to contrast the omnidirectional spin Hall effect
described by Eq.~(\ref{eq:delta-J_spin-1}) to the usual spin Hall effect due
to Rashba SOC acting in the $xy$ plane \cite{SinovaPRL}.  ln the case
of the Rashba SOC,
the spin Hall  response to an external force can be presented in a
manner similar to Eq.~(\ref{eq:delta-J_spin-1}). Specifically, the spin
current resulting from driving the system along $\bs e_\beta$ can be written
as 
\be
(\delta J_\text{spin}^\text{Rashba})^z_i 
\sim \bs e_z \cdot (\bs e_i \times \bs e_\beta)\,,
\label{eq:delta-J_spin-Rashba}
\ee
so the induced spin current carries only the $z$
component of the spin, which is perpendicular to the  SOC
plane  ($xy$).  The spin  Hall current  given by Eq.~(\ref{eq:delta-J_spin-Rashba}) is zero if the driving direction
$\bs e_\beta$ or the direction   $e_i$ of the induced spin current are
 taken to be along 
$\bs e_z$. 
 In the case of the Weyl SOC, the induced spin current  given by Eq.~(\ref{eq:delta-J_spin-1})  carries spins  polarized
in any direction $\bs e_j$.   The  induced spin current and the driving
can then  point in  arbitrary directions  $\bs e_i$ and $\bs e_{\beta}$ as long as they are not parallel to each other.

\subsection{Momentum averaging\label{subsec:averaging}} 

Although we are dealing with a 3D system of atoms, it is convenient to define a
generic $D$-dimensional particle density function $n_{D}$ for a chemical
potential $\mu$ at a temperature $T$:
\be
n_{D} =n_{D+} + n_{D-}\,,
\ee
where
\be
n_{D\pm} = \frac{S_D}{(2\pi)^D} \int_0^\infty p^{D-1} dp \, f^{\pm}(p)\,
\ee
is a $D$-dimensional density of atoms in the upper or lower dispersion branch,
$ S_D\equiv 2\pi^{D/2} / \Gamma(D/2) $ is  a $D$-dimensional unit-sphere area,
\be
  f^\pm(p) = [e^{(\ve_{p\pm}  -\mu)/k_B T} + \alpha]^{-1}
 \label{eq:f^pm}
\ee
is  a distribution function for fermions ($\alpha = 1$) or bosons ($\alpha =
-1$) in the dispersion branch $\ve_{p\pm}$, and $k_B$ is the Boltzmann
constant.

We consider a system,  with a fixed $3D$ particle density $\nu$.  The chemical
potential $\mu$ at a certain temperature $T$ is obtained from the condition
\be
n_{3}(\mu,T,\chi)=\nu.
\ee
Using this notation, the spin Hall conductivity (\ref{eq:sigma_SHE-result})
takes the form 
\be
\sigma_{\text{SH}}
=
\frac{ n_{2-}-n_{2+} }{6\chi \pi}
 \,,
\label{eq:spin-current-T}
\ee
where $n_{2+}$ and $n_{2-}$ correspond to the 2D densities of atoms in the
upper and lower dispersion branches, respectively.

In particular, an ensemble of fermions with a thermal energy $k_B T$ much
smaller that the Fermi energy $\ve_F$ populate the energy levels up to
$\ve_F=\mu$ corresponding to the zero temperature limit. If $\ve_F$ is below
the band crossing, only the lowest dispersion band is populated ($n_{2+}=0$).
On the other hand, if $\ve_F$ is above the band crossing, both dispersion band
are populated.  In both cases the difference in band densities is given by  
\be
n_{2-}-n_{2+}= \frac{\chi}{\pi}\sqrt{2\mu m}\,.
\label{eq:n_pm-all}
\ee
Using Eqs.~(\ref{eq:spin-current-T}) and (\ref{eq:n_pm-all}) one can see that
for fermions  the spin Hall conductivity  $\sigma_{\text{SH}}\propto
\sqrt{\mu}$  depends on the SOC strength $\chi$ only through the chemical
potential  $\mu=\mu\left(\chi,\nu \right)$ in the zero temperature limit.
This differs from the previously considered 2D Rashba SOC where
the low temperature spin Hall conductivity $\sigma_{\text{SH}}^{\text{(2D)}}$
takes a universal value which is independent of the SOC strength if both
bands are populated \cite{SinovaPRL}.

\begin{figure}[t]
\includegraphics[width=0.49\linewidth]{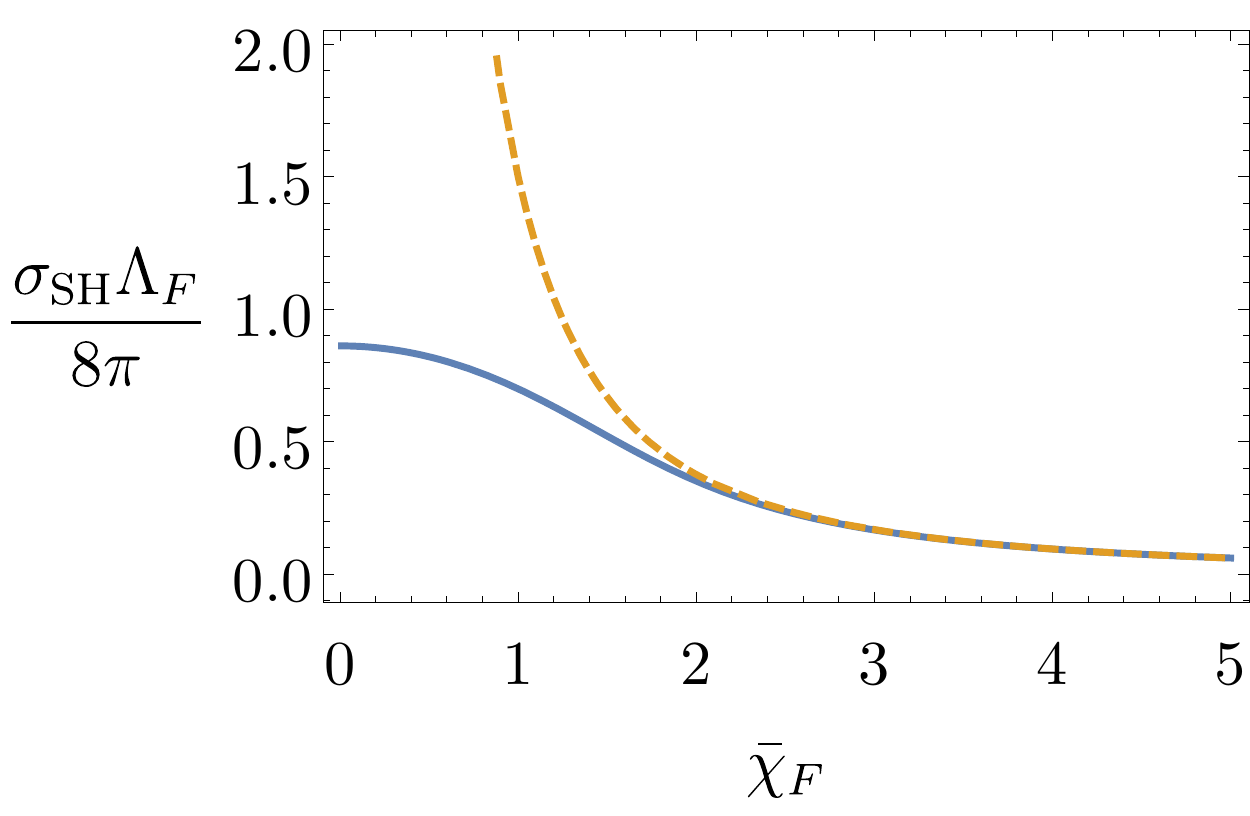}
\includegraphics[width=0.49\linewidth]{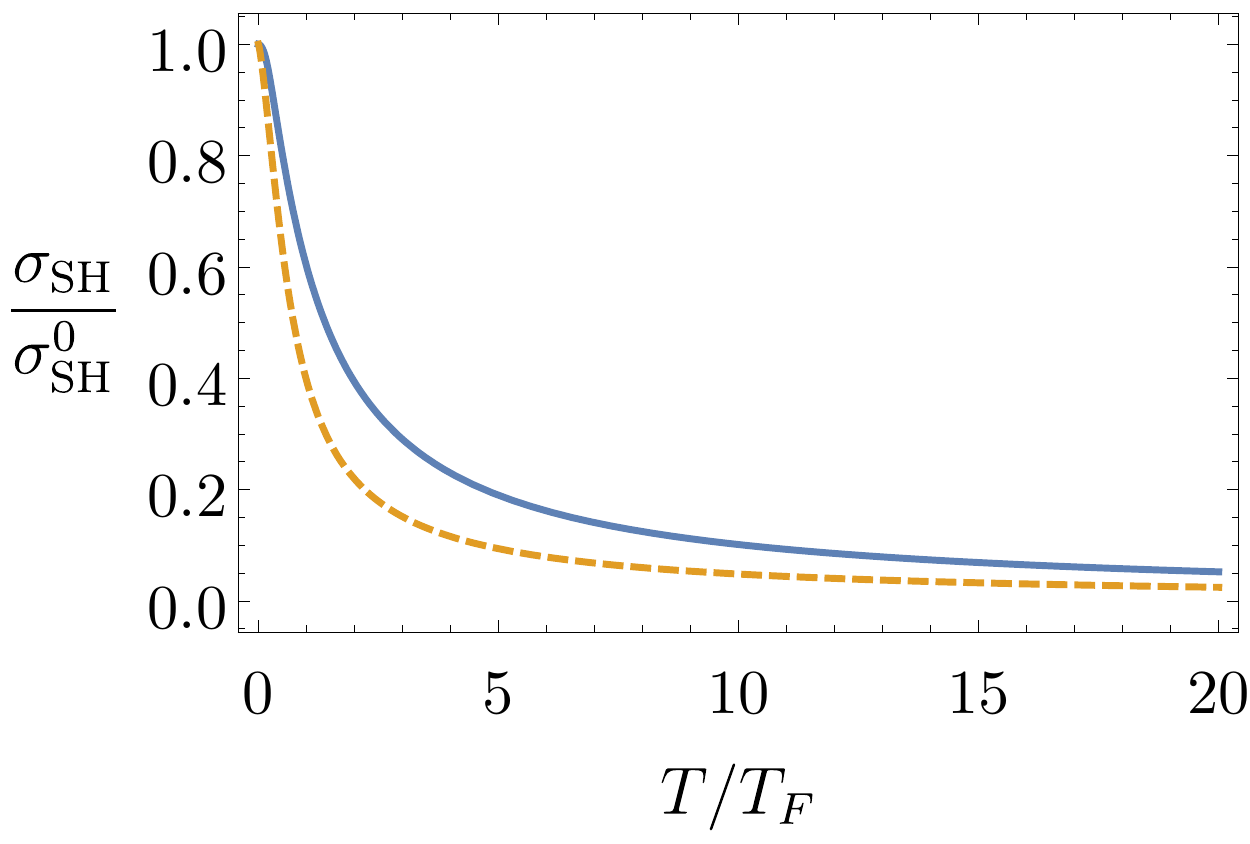}
\caption{Left: dimensionless spin Hall conductivity as a function of the 
  dimensionless SOC strength.  
  Right: temperature dependence of spin Hall conductivity for a
  fixed SOC strength $\bar \chi_F = 1$ normalized to spin Hall conductivity at
  low temperature. Both plots show the results for fermions (blue solid) and
  bosons (yellow dashed) in the absence of a condensate.}
\label{fig:sigma-fermi-bose-low-t}
\end{figure}

In general the spin-Hall conductivity  $\sigma_{\text{SH}}$ depends  on the
temperature, the statistical distribution, and the SOC strength.  We explore
these dependencies in Fig.~\ref{fig:sigma-fermi-bose-low-t},  in which the spin
Hall conductivity is plotted for  the fixed particle density $\nu$ as a
function of the temperature and the dimensionless SOC strength
\be
\bar\chi_F =\chi\sqrt{2 \beta_F/m}\,,
\ee
where $\beta_F = 1/k_B T_F$  and 
\be
T_F \equiv (3\pi^2 n)^{2/3}/{2mk_B}
\ee
is defined in the same way both  for bosons and fermions. For fermions $T_F$
corresponds to the Fermi temperature.  In addition we define the de~Broglie
wavelength at the temperature $T_F$
\be
\Lambda_F = \sqrt{2\pi\beta_F/m} = (16/\pi n^2)^{1/6}\,.
\ee

The proposed effect is present both for bosons and fermions.  Even though the
induced conductivity is the largest for $\bar \chi_F \rightarrow 0$, care must
be taken in interpreting this result.  In fact, in this parameter range the
adiabatic approximation becomes invalid,  as it will be discussed in detail in
the next Subsection.    Note that at a mean-field level, the conductivity would
not be modified by the presence of interactions, as they would merely shift the
chemical potential by a constant.   Yet considering a Bose-Einstein condensed
state in this system is inherently nontrivial due to the absence of a single
minimum in the dispersion \cite{Stanescu2008,SOCreview14,Zhai14-review}.   Even
small interactions will have a large effect on the nature of the condensate
groundstate, and, in turn, on its transport properties.  Hence, the results
presented here only hold for noncondensed gases with weak interactions. 

\subsection{Validity of approximation\label{subsec:Validity}} 

In our analysis we have applied the adiabatic approximation which is applicable
when Eq.~\eqref{eq:adiabatic_condition} holds.  Here we explicitly check if
this approximation holds for a typical experimental system in the range of
interest of the parameters.  Given a sufficiently low temperature and particle
density, when only the bottom of the lowest band is occupied, we can assume
that $|\bs p| \simeq \chi$.  Moreover, we assume for clarity that the driving
is provided by some harmonic potential with the displacement of the center of
the system equal to the length of the trap.  In that case
Eq.~\eqref{eq:adiabatic_condition} immediately yields the condition
$\omega\beta_F \ll \bar \chi_F^2$, where $\omega$ is the trap frequency.  We
consider a system with  a particle density of $10^{18}$m$^{-3}$, which
corresponds to $T_F \simeq 20$nK. For these parameter values, a driving force
provided by a $\omega=10$Hz trap leads to the adiabaticity criterion $\bar
\chi_F \gg 0.1$.

Hence, our approximation is certainly valid in a setting when driving is
relatively gentle, temperature is very low, and SOC strength is moderate.  This
regime, where $\sigma_\text{SH}$ is maximized and the approximation is robust,
does not seem to put any extra challenges to the experimentalist, besides
achieving the Weyl SOC.  We note that temperatures as low as several nanokelvin
have been demonstrated \cite{Medley11}, while an optically-generated SOC
routinely achieves $\bar\chi_F = 5$ in the equal Rashba-Dresselhaus case
\cite{Beeler13}.  The question of validity of the adiabatic approximation is,
however, separate from the feasibility of detecting this effect. The latter
question is addressed in the next subsection.

In applying our adiabatic approximation we have implicity assumed that the
driving is switched on slowly.  However, if the driving is switched on
suddenly, the adiabatic approximation is not sufficient, and one has to solve
the LLE at least to the second order in time derivatives.  We have checked,
however, that the SHE is still present in this post-adiabatic solution. The
only new feature that arises in this higher-order solution is a
Zitterbewegung-like beating between the two adiabatic solutions, which has been
considered elsewhere
\cite{Vaishnav08PRL,Merkl08,Engels2013,LeBlanc2013,Zulicke07}.

\subsection{Detection of spin current\label{subsec:Detection}}

As discussed above, the most direct signature of the omnidirectional Hall
effect is the spin current $\delta J_\text{spin}$. The experimental sequence
needed to detect that current depends on the precise details of the
implementation of Weyl SOC, as proposals to achieve it utilize qualitatively
different physical means \cite{3dSOCprl,magSOCprl,Brandon13}.  Nevertheless,
several general remarks can be made with no reference to these experimental
details.

In particular, it is possible to utilize the fact that $\delta J_\text{spin}$
given by Eq.~(\ref{eq:delta-J_spin-1}) is  the transverse spin current. This is
beneficial, since the equilibrium spin current $J^0_\text{spin}$ is
longitudinal, and the spin of an atom in the upper (lower) band points along
(opposite to) the  momentum  $\bs p $.  As a result, one can filter out the
$\delta J_\text{spin}$ signal by choosing a beneficial configuration of the
driving direction $\bs e_\beta$, the spin (Stern-Gerlach) projection  axis $\bs
e_j=\bs e_\text{SG}$ and the direction of the momentum measurement $\bs e_j=\bs
e_\text{det}$.   Specifically, if one  takes these three  vectors to be
orthogonal, the triple product
\be
\bs e_\text{det} \cdot (\bs e_\text{SG} \times \bs e_\beta)
\ee
featured in Eq.~(\ref{eq:delta-J_spin-1}) for  $\delta J_\text{spin}$ is
maximized and the signal is the strongest.  A relation between the spin current
and the spin (Stern-Gerlach) projection measurement is considered in the last
subsection of the Appendix.

Moreover, one can estimate the size of the effect of the omnidirectional spin
Hall effect on the momentum distribution. Since the SOC strength $\chi$ sets
the characteristic momentum in the distribution of particles in the system, the
magnitude of the signal (the change of the momentum distribution due to the
omnidirectional spin Hall effect) is approximately equal to the ratio
$\dot\beta \sigma_\text{SH}/\nu\chi$.

\subsection{Spin-current induced by a time-dependent Zeeman term \label{subsec:Zeeman}}

The spin Hall effect can also be induced by a time-dependent Zeeman shift
rather than by a time-dependent external force. In that case the term $\gamma
\bs e_\gamma \cdot \hat{\bs \sigma}$ is to be added to the Hamiltonian, and the
effective magnetic field $\bs M$ determining the spin dynamics becomes 
\be
\bs M = \chi\bs p /m + \gamma \bs e_\gamma \,.
\label{eq:M-time-Zeeman}
\ee
Since the scalar driving $\beta$  (due to a spin-independent force on an atom),
and the Zeeman driving $\gamma$  (due to a magnetic pulse) enter the effective
magnetic field $\bs M$ in the same manner, these two ways of driving the system
lead to the same effect for the spin dynamics. Therefore, the above analysis of
the induced spin current due to the spin-independent force can be transferred
in a straightforward manner to the case of the Zeeman driving via the
replacement of $\beta $ by $-m \gamma $ and $\bs e_\beta$ by $\bs e_\gamma$. 

\section{Summary and future work\label{sec:Summary}}  

In summary, we have put forward a proposal to observe inherently non-Abelian
dynamics in the form of an omnidirectional spin Hall effect in a driven system
in the presence of a Weyl (three-dimensional) spin-orbit coupling. We have
discussed two independent  ways  to drive the system leading to the same effect
for the spin dynamics: either through a constant spin-independent force or a
time-dependent Zeeman field. We have also evaluated the strength of this effect
in terms of conductivity for noninteracting uncondensed bosonic or fermionic
gas.  All of the components of this proposal seem to be within the reach of
cold-atom experiments in the near future, and their combination has a potential
to unambiguously demonstrate non-Abelian dynamics in a continuum  (non-lattice)
cold-atom system for the first time.

In future work, we plan to investigate collective modes of a trapped system and
look for signatures of the non-Abelian dynamics described here. Other promising
avenues of research include a more careful account of interactions, especially
with the Bose-Einstein condensation in mind, and also considering the kinetic
effects in this system, e.g., the relaxation of spin current also known as spin
drag \cite{spindragExp}, which was not considered here. 

\begin{acknowledgments}
It is our pleasure to thank 
Egidijus Anisimovas,
Alain Aspect, 
Denis Boiron, 
Rembert Duine,
Gabriele Ferrari,
Lars Fritz,
Simonas Grubinskas,
Krzysztof Jachymski,
Wolfgang Ketterle,
Oleksandr Marchukov,
Bill Phillips, 
Henk Stoof,
Roland Winkler,
and 
Ulrich Zuelicke
for stimulating discussions.
J.~A. has received funding from the European Union's Horizon 2020 research and
innovation programme under the Marie Sk{\l}odowska-Curie grant agreement
No~706839 (SPINSOCS).
\end{acknowledgments}

\appendix*

\section{Spin current\label{sec:appendix-spin-current}}

In this Appendix we motivate the definition of the spin current given in the
text by deriving the spin continuity equation and considering the effect of a
spin (Stern-Gerlach) projection on the velocity operator.  In contrast to the
main text, in this Appendix we use hats to label not only the spin operators
but all operators (including the coordinate and momentum operators
$\hat{\boldsymbol{r}}$ and $\hat{\boldsymbol{p}}$) in order to make the
Appendix as accessible as possible.

\subsection{Continuity equation and spin current}

The spin density is a vector field given by
\begin{equation}
\boldsymbol{\rho}(\boldsymbol{r})=
\Psi^{\dag}(\boldsymbol{r})\hat{\boldsymbol{\sigma}}\Psi(\boldsymbol{r})
=\langle \Psi | \boldsymbol{r}\rangle \hat{\boldsymbol{\sigma}}
\langle\boldsymbol{r}|\Psi\rangle
=\langle \Psi |\hat{\boldsymbol{\rho}}_{\boldsymbol{r}}|\Psi\rangle\,,
\label{eq:rho}
\end{equation}
where $\hat{\boldsymbol{\rho}}_{\boldsymbol{r}}=
\hat{\boldsymbol{\sigma}}\delta(\hat{\boldsymbol{r}}-\boldsymbol{r})$ is the
corresponing spin density operator, and
$\Psi(\boldsymbol{r})=\langle\boldsymbol{r}|\Psi\rangle$ is a two component
column-spinor.  Here the quantum average has been carried out over the full
state-vector $|\Psi\rangle$  accommodating both the motional and spin degrees
of the atom.  Furthermore, we have casted the operator
$\delta(\hat{\boldsymbol{r}}-\boldsymbol{r})=|\boldsymbol{r}\rangle\langle\boldsymbol{r}|$
in terms of the eigenstates $|\boldsymbol{r}\rangle$ of the coordinate
operator: $\hat{\boldsymbol{r}}|\boldsymbol{r}\rangle=\bs
r|\boldsymbol{r}\rangle$. 

The dynamics of the operator $\hat{\boldsymbol{\rho}}_{\boldsymbol{r}}$ is
governed by the Hamiltonian  given by Eq.~\eqref{eq:H} which contains the Weyl
SOC term, and thus
\begin{align}
\frac{d}{dt}\hat{\boldsymbol{\rho}}_{\boldsymbol{r}}=&
\frac{1}{i}[\hat{\boldsymbol{\rho}}_{\boldsymbol{r}},\hat{H}]
\\\nonumber
=&\frac{1}{2}\sum_l\left(\frac{1}{i}[\hat{\boldsymbol{\rho}}_{\boldsymbol{r}},
\hat{p}_{l}-\hat{\mathcal{A}}_{l}]\hat{v}_{l}
+\hat{v}_{l}\frac{1}{i}[\hat{\boldsymbol{\rho}}_{\boldsymbol{r}},
\hat{p}_{l}-\hat{\mathcal{A}}_{l}]\right)\,,\label{eq:tmp-1}
\end{align}
where the matrix-valued velocity operator $\hat{\boldsymbol{v}}$ is defined in
Eq.~(\ref{v-definition}). Since
\begin{equation*}
[\hat{\boldsymbol{\rho}}_{\boldsymbol{r}},\hat{p}_{l}-\hat{\mathcal{A}}_{l}]=
-i\boldsymbol{\sigma}\nabla_{l}\delta(\hat{\boldsymbol{r}}-\boldsymbol{r})
-[\boldsymbol{\sigma},\hat{\mathcal{A}}_{l}]\delta(\hat{\boldsymbol{r}}-\boldsymbol{r})\,,
\end{equation*}
one arrives at the following continuity equation: 
\begin{equation}
\frac{d}{dt}\hat{\boldsymbol{\rho}}_{\boldsymbol{r}}
+\nabla_{l}\hat{\boldsymbol{j}}_{\boldsymbol{r}l}
=\hat{\mathbf{G}}_{\boldsymbol{r}}\,,
\end{equation}
where
\begin{equation}
\hat{\mathbf{G}}_{\boldsymbol{r}}=
-\frac{1}{2i}
\left(\delta(\hat{\boldsymbol{r}}
-\boldsymbol{r})[\hat{\boldsymbol{\sigma}},\hat{\mathcal{A}}_{l}]\hat{v}_{l}
+\hat{v}_{l}[\hat{\boldsymbol{\sigma}},\hat{\mathcal{A}}_{l}]\delta(\hat{\boldsymbol{r}}
-\boldsymbol{r})\right)
\label{eq:G_r}
\end{equation}
is the spin source operator and
\be
\hat{\boldsymbol{j}}_{\boldsymbol{r}l}=\frac{1}{2}(\delta(\hat{\boldsymbol{r}}
-\boldsymbol{r})\hat{\boldsymbol{\sigma}}\hat{v}_{l}
+\hat{v}_{l}\hat{\boldsymbol{\sigma}}\delta(\hat{\boldsymbol{r}}
-\boldsymbol{r}))
\label{eq:j_r}
\ee
is the probability current operator.

In what follows we shall consider the spin current for momentum eigenstates
of the Weyl SOC Hamiltonian,
\begin{equation}
\Psi(\boldsymbol{r})=\Psi_{\mathbf{p}\pm}(\boldsymbol{r})=
V^{-1/2}\eta_{\mathbf{p}\pm}e^{i\mathbf{p}\cdot\mathbf{r}}\,,
\label{eq:eigenstates}
\end{equation}
where $V$ is a quantization volume and the spinor $\eta_{\mathbf{p}\pm}$
describes the quantum states for the spin along or opposite to the momentum:
$\mathbf{p}/p\cdot\hat{\boldsymbol{\sigma}}\eta_{\mathbf{p}\pm}=
\mp\eta_{\mathbf{p}\pm}$. The corresponding expectation value of
the spin current is
\begin{equation}
\boldsymbol{j}_{l}(\boldsymbol{r})=\left\langle \Psi\right|
\hat{\boldsymbol{j}}_{\boldsymbol{r}l}|\Psi\rangle=\frac{1}{2V}
\langle\{ \hat{\boldsymbol{\sigma}},
\hat{v}_{l}\} \rangle\,,\label{eq:j_r--quant-average}
\end{equation}
with $\{ \hat{\boldsymbol{\sigma}},\hat{v}_{l}\}
=\hat{\boldsymbol{\sigma}}\hat{v}_{l}
+\hat{v}_{l}\hat{\boldsymbol{\sigma}}$,
where the brackets $\langle\cdots\rangle$ label the quantum averaging over
the spinor state $\eta_{\mathbf{p}\pm}$. Performing also a statistical
averaging over the atomic single-particle distribution $f^{\pm}(\mathbf{p})$, 
one arrives at the spin current presented in Eq.~\eqref{eq:J_spin} of the main text:
\begin{equation}
(J_{\text{spin}})_{i}^{j}=\sum_{\boldsymbol{p},\eta=\pm} \frac{ f^{\eta}(\boldsymbol{p})}{2V}
\langle \{ \hat{\sigma}_{j},\hat{v}_{i} \} \rangle
=\frac{1}{2}\overline{\langle\{ \hat{\sigma}_{j},\hat{v}_{i}\} \rangle}
\,,
\label{eq:j_r--quant-average-1}
\end{equation}
where the statistical averaging is denoted by an overline.

\subsection{Source term}

The vector potential $\hat{\mathbfcal{A}}$ given by Eq.~\eqref{eq:H} describes
the 3D SOC and the driving. The space- and spin-independent driving term
$\beta(t)\mathbf{e}_{\beta}$ does not contribute to the commutators entering
Eq.~(\ref{eq:G_r}), giving
\be
[\hat{\sigma}_{j},\hat{\mathcal{A}}_{l}]\hat{v}_{l}
=\chi[\hat{\sigma}_{j},\hat{\sigma}_{l}]\hat{v}_{l}
=2i\chi\varepsilon_{jln}\hat{\sigma}_{n}\hat{v}_{l}
=-2i\chi(\boldsymbol{\hat{\sigma}}\times\hat{\boldsymbol{v}})_{j}\,.
\ee
In a similar way,
\begin{equation}
\hat{v}_{l}[\hat{\sigma}_{j},\hat{\mathcal{A}}_{l}]
=2i\chi(\hat{\boldsymbol{v}}\times\boldsymbol{\hat{\sigma}})_{j}\,.
\end{equation}
Consequently, 
\begin{equation}
\hat{\mathbf{G}}_{\boldsymbol{r}}=\chi\left(\delta(\hat{\boldsymbol{r}}
-\boldsymbol{r})\hat{\boldsymbol{\sigma}}\times \hat{\boldsymbol{v}}
-\hat{\boldsymbol{v}}\times\hat{\boldsymbol{\sigma}}\delta(\hat{\boldsymbol{r}}
-\boldsymbol{r})\right)\,.
\end{equation}
with $m\hat{\boldsymbol{v}}=\hat{\boldsymbol{p}}-\chi\hat{\boldsymbol{\sigma}}
-\beta\mathbf{e}_{\beta}$.

In the case of the 3D SOC, the eigenstates $\eta_{\mathbf{p}\pm}$ describe the
spin along the momentum:  $\langle \hat{\boldsymbol{\sigma}}\rangle =\pm
\mathbf{p}/p$ .  Therefore, the source term vanishes after taking the quantum
expectation value $\langle \hat{\mathbf{G}}_{\boldsymbol{r}} \rangle $ and
averaging over an isotropic momentum distribution.

\subsection{Spin projection measurement}

Here we consider the effect of the spin (Stern-Gerlach) projection measurement
on the velocity $\hat{v}_{i}$  along the unit vector $\bs e_i$.  We will show
that the spin current measured in this way is consistent with its previous
definition. In particular, a Stern-Gerlach (SG) projection in the $j$
direction is given by the projector $ |s_{j}\rangle\langle s_{j}| $, where the
quantum state $|s_{j}\rangle$ describes the spin pointing along the unit vector
$\bs e_j$. Calculating the expectation value of the velocity operator
$\hat{v}_{i}$ with respect to such spin-projected states entails evaluating
$|s_{j}\rangle\langle s_{j}|\hat{v}_{i}|s_{j}\rangle\langle s_{j}|$. Since the
spin projection operator can be written down as 
\be
\hat{I}_{s_{j}}=|s_{j}\rangle\langle s_{j}|=\frac{1}{2}(1+\bs e_{j}\cdot\hat{\bs \sigma}),
\ee
we have 
\begin{multline}
\hat{I}_{s_{j}}
\hat{v}_{i}
\hat{I}_{s_{j}} =  
\big(
\hat{v}_{i}+
\left(\bs e_{j}\cdot\hat{\bs \sigma}\right)
\hat{v}_{i}
\left(\bs e_{j}\cdot\hat{\bs \sigma}\right)
\big)/4
\\
+
\big(
\left(\bs e_{j}\cdot\hat{\bs \sigma}\right) \hat{v}_{i}
 +  
\hat{v}_{i}\left(\bs e_{j}\cdot\hat{\bs \sigma}\right)\big)/4.
\end{multline}
Comparing this expression with Eq.~\eqref{eq:J_spin},  one can see that the
second term on the right-hand side is proportional to the spin current. Using
the properties of Pauli matrices, the first term simplifies to
\begin{multline}
\hat{v}_{i}
+
\left(\bs e_{j}\cdot\hat{\bs \sigma}\right)\hat{v}_{i}
\left(\bs e_{j}\cdot\hat{\bs \sigma}\right)
\\
=2\left[p_{i}-\chi\left(\bs e_{j}\cdot\hat{\bs \sigma}\right)
\left(\bs e_{i}\cdot \bs e_{j}\right)\right]/m.
\end{multline}
 Consequently 
\be
\hat{I}_{s_{j}}
\hat{v}_{i}
\hat{I}_{s_{j}}  
= \frac{p_{i}-\chi\left(\bs e_{j}\cdot\hat{\bs \sigma}\right)
\left(\bs e_{i}\cdot \bs e_{j}\right)}{2m}
+
\frac{(\hat{J}_{\text{spin}})_{i}^{j}}{2}\,.
\ee
As the projection direction is reversed,  $\bs e_j \rightarrow
-\bs e_j$, the first term is unaffected, while the second term changes  its
sign. Therefore by considering  the difference in velocities between the spin
up and the spin down components resulting from a spin (Stern-Gerlach)
projection in  the direction $\bs e_j$,
one measures the spin current exactly as defined in Eq.~\eqref{eq:J_spin} in
the main text.

\bibliography{omnidirectional}

\end{document}